\documentclass[aps,prd,preprintnumbers,nofootinbib]{revtex4} 
\usepackage{epsfig}

\def\Str{\textrm{Str}}

\def\CA{{\cal A}}

\def\CD{{\cal D}}

\def\CL{{\cal L}}

\def\CS{{\cal S}}

\def\spose#1{\hbox to 0pt{#1\hss}}
\def\ltapprox{\mathrel{\spose{\lower 3pt\hbox{$\mathchar"218$}}
 \raise 2.0pt\hbox{$\mathchar"13C$}}}
\def\gtapprox{\mathrel{\spose{\lower 3pt\hbox{$\mathchar"218$}}
 \raise 2.0pt\hbox{$\mathchar"13E$}}}
\def\inapprox{\mathrel{\spose{\lower 3pt\hbox{$\mathchar"218$}}
 \raise 2.0pt\hbox{$\mathchar"232$}}}

\preprint{UW/PT 03-11}
\begin{document}

\title{Mass dependence of the hairpin vertex in quenched QCD}

\author{Stephen R. Sharpe}
\email{sharpe@phys.washington.edu}
\affiliation{Physics Department, University of Washington,
Seattle, WA 98195-1560}


\begin{abstract}
The pseudoscalar ``hairpin'' vertex
(i.e. quark-disconnected vertex) plays a key role
in quenched chiral perturbation theory. Direct calculations
using lattice simulations find that it has
a significant dependence on quark mass.
I show that this mass dependence can be used to determine
the quenched Gasser-Leutwyler constant $L_5$.
This complements the calculation of $L_5$ using the mass dependence of
the axial decay constant of the pion.
In an appendix, I discuss power counting for quenched chiral perturbation
theory and describe the particular scheme used in this paper.
\end{abstract}

\pacs{12.38.Gc, 11.15.Ha, 11.30.Rd, 12.39.Fe}

\maketitle

\section{Introduction}
\label{sec:intro}

Although the quenched approximation to QCD is 
being gradually superseded by unquenched simulations,
it remains useful as a qualitative
guide to the physics of QCD, and as a testing
ground for new methods of discretizing light fermions.
In the latter role, the unphysical chiral singularities of quenched QCD
turn out to be helpful---reproducing them provides evidence
that the light fermion method under study can approach the chiral limit.
Examples of these singularities are the enhanced chiral logarithms in
the pion mass and the pseudoscalar decay constant~\cite{BG,SSChlog}.
The source of these singularities is the
hairpin vertex, i.e. the appearance of a double-pole contribution
in the flavor-singlet pion two-point function.
In infinite volume (where the contributions of exact zero modes
vanish) this vertex is the key unphysical feature of
quenched QCD.

Quenched chiral perturbation theory (Q$\chi$PT) allows one to
systematically predict the form of these singularities.
Although Q$\chi$PT does not have the same
theoretical underpinnings as chiral perturbation theory 
for unquenched QCD, and thus must be treated as a phenomenological
theory, it has been quite successful in describing the properties
of light pseudo-Goldstone bosons.
A particularly thorough study has been undertaken 
by Bardeen {\em et al.}~\cite{BDET,BET} 
(referred to as BDET in the following).\footnote{%
See also the work of 
Refs.~\cite{Kuramashi,Heitger,DeGrand,Liu,Chiu,Gattringer}.}
They find that consistent values of
the hairpin vertex can be obtained
both from a direct calculation of the hairpin correlator and from
fitting to the chiral logarithms predicted from loop contributions
to various quantities.

There is, however, a puzzling aspect to the results 
for the hairpin vertex.
On the one hand, the vertex is found to have substantial dependence
on quark mass. As I argued in Ref.~\cite{SSLAT96},
this can be explained by a non-zero value for the momentum-dependent
part of the hairpin ($\alpha_\Phi\sim 0.5$ in the notation 
explained below).
On the other hand, BDET note that, at tree-level, 
one can also determine $\alpha_\Phi$ from the
coefficient of the 
sub-leading single-pole contribution to the hairpin correlator.
They find that $\alpha_\Phi$ determined in this way is much
smaller than that determined from the mass dependence,
and, in fact, is consistent with zero. 

The purpose of this note is to resolve the discrepancy between
these two methods for determining $\alpha_\Phi$. It turns out that,
at next-to-leading-order (NLO) in quenched chiral perturbation theory,
the hairpin vertex picks up a mass dependence proportional to
$L_5$ (the quenched analog of corresponding 
Gasser-Leutwyler constant), in addition to that proportional
to $\alpha_\Phi$. The coefficient of the single pole, however, is
not affected at this order, and remains
proportional to $\alpha_\Phi$ alone. 
This allows a consistent description of the hairpin results:
one should use the BDET method to determine $\alpha_\Phi$,
with a result close to zero, while the mass dependence of
the hairpin vertex determines $L_5$.
This provides a check of determinations of $L_5$
based on the mass dependence of the axial decay constant.

\section{Details of Calculation}
\label{sec:details}

I consider quenched QCD with $N_V$ valence quarks.
For the following considerations I need $N_V\ge 2$, and for
definiteness I use $N_V=3$.
I take all the quarks to be degenerate, since this is what has
been done in numerical evaluations of the hairpin vertex.
The leading order quenched chiral Lagrangian is then~\cite{BG,SSChlog}
\begin{equation}
\CL_0 = \frac{f^2}{4} \Str\left(
\partial_\mu U \partial_\mu U^\dagger
-\chi U - U^\dagger \chi\right)
+ m_0^2 \Phi_0^2 + \alpha_\Phi (\partial_\mu \Phi_0)^2
\,,
\label{eq:L0}
\end{equation}
where ``$\Str$'' indicates supertrace,
$U=\exp(2 i \Phi/f)$ is an element of $U(3|3)$, 
$\Phi$ contains the pseudo-Goldstone
bosons and fermions, and $\Phi_0=\Str(\Phi)/\sqrt3$
is the ``super-$\eta'$'' field.
Quark masses are contained within  $\chi=2 B_0 M$,
where $M=\textrm{diag}(m,m,m,m,m,m)$ is the mass matrix of quarks and ghost-quarks.
I assume that $m$ is small enough that it is appropriate
to use chiral perturbation theory to describe the properties
of the pseudo-Goldstone hadrons.

The parameters $f$ (pion axial decay constant in the chiral limit, normalized
so it is close to $90\;$MeV) and $B_0$
are the quenched analogs of the corresponding parameters that appear in the
unquenched chiral Lagrangian. 
The parameters $m_0$ and $\alpha_\Phi$ are, however, special
to quenched chiral perturbation theory, 
in which one cannot integrate out the $\Phi_0$. 
In the large $N_c$ and chiral limits, one can show that
$m_0=m_{\eta'}$ \cite{BG,SSChlog}.
Assuming this holds also at $N_c=3$, one expects $m_0\approx 0.85\;$GeV.
The presence of the new mass-scale $m_0$ gives rise to the enhanced
singularities of the quenched theory, the strength of which
is parameterized by $\delta = m_0^2/(48 \pi^2 f^2)$. 
It will be useful in the following to keep in mind the approximate
value of $\delta$.
The naive estimate
(using QCD values for the parameters) is $\delta\approx 0.2$,
while simulations find values in the range $\delta=0.1-0.2$
(for a recent review see Ref.~\cite{Wittiglat2002}).
What is most important about these values is that they are small.

The present calculation will be performed at 
NLO in a power counting scheme for quenched chiral perturbation theory
described in the appendix.
It will require the standard NLO Lagrangian of the unquenched theory,
suitably generalized by changing traces to supertraces, plus
an additional term:
\begin{eqnarray}
\CL_{1A} &=& 
-L_1 \left[\Str\left(\partial_\mu U \partial_\mu U^\dagger\right)\right]^2
-L_2\, \Str\left(\partial_\mu U \partial_\nu U^\dagger\right)
\Str\left(\partial_\mu U \partial_\nu U^\dagger\right)
-L_3\, \Str\left(\partial_\mu U \partial_\mu U^\dagger
\partial_\nu U \partial_\nu U^\dagger\right) 
\nonumber \\
&&\mbox{}
+ L_4\, \Str\left(\partial U \partial U^\dagger\right)
\Str\left(\chi U^\dagger + U \chi\right)
+ L_5\, \Str\left(\partial U \partial U^\dagger
\left[\chi U^\dagger + U \chi\right]\right)
- L_6 \left[\Str\left(\chi U^\dagger + U \chi\right)\right]^2
\nonumber \\
&&\mbox{}
- L_7 \left[\Str\left(\chi U^\dagger - U \chi\right)\right]^2
- L_8 \left\{
\Str\left(U\chi U\chi\right) + 
\Str\left(\chi U^\dagger \chi U^\dagger\right)
 \right\} 
\nonumber \\
&&\mbox{}
-L_{Q}\, \Str\left(\partial_\mu U \partial_\nu U^\dagger
\partial_\mu U \partial_\nu U^\dagger\right)
\,.\label{eq:L1A}
\end{eqnarray}
Here $L_{1-8}$ are the quenched analogs of the corresponding
Gasser-Leutwyler coefficients. $L_Q$ multiplies an operator
which is not independent in unquenched $SU(2)$ or $SU(3)$ chiral
perturbation theory, but which is independent in the quenched
and partially quenched theories. Its presence in partially quenched
theories was noted and explained recently~\cite{ruth}. 
That it is also needed in the quenched
theory can be seen by generalizing the arguments given there.

In the quenched theory, one must also include other operators
obtained from those in $\CL_0$ and $\CL_{1A}$ by multiplying by
powers of $\Phi_0$. These introduce many new constants.
It turns out, however, that the resulting operators do not
contribute at NLO to the quantities I consider here,
as explained in the appendix.
For illustration, I list
two of the leading operators of this type:
\begin{eqnarray}
\CL_{1B}&=& 
v_0 (m_0^2/f^2) \Phi_0^4
- v_8 L_8 (i\Phi_0/f)\left\{
\Str\left(U\chi U\chi\right) - 
\Str\left(\chi U^\dagger \chi U^\dagger\right)
		\right\} 
 \,,\label{eq:L1B} 
\end{eqnarray}
where $v_{0,8}$ are additional couplings.
\bigskip

Following BDET, I consider the following correlation functions:
\begin{eqnarray}
\widetilde\Delta_c(p) &=& 
\int_x e^{-ip\cdot x} 
\langle \bar q_A \gamma_5 q_B(x) \ \bar q_B\gamma_5 q_A(0)\rangle
\,,\label{eq:PPconn}
\\
\widetilde \Delta_h(p) &=& 
\int_x e^{-ip\cdot x} 
\langle \bar q_A \gamma_5 q_A(x) \ \bar q_B\gamma_5 q_B(0)\rangle
\,.\label{eq:PPdisc}
\end{eqnarray}
Here $q_A$ and $q_B$ are two different valence quarks, 
and the expectation value
indicates the quenched average. 
The subscripts $c$ and $h$ are a reminder that
only quark-connected contractions contribute 
to $\widetilde\Delta_c$, whereas $\widetilde\Delta_h$ contains only
quark-disconnected contractions.
In other words, $\widetilde\Delta_c$ is the
usual non-singlet pseudoscalar two-point correlator, 
while $\widetilde\Delta_h$ is the pseudoscalar hairpin correlator. 
These are the correlators used in simulations to study the
hairpin vertex.

At low momenta the connected correlator is well described by
a single pole
\begin{equation}
\widetilde\Delta_c(p) = - \frac{f_P^2}{p^2 + M_\pi^2} + \textrm{non-pole}
\,,
\label{eq:singlepole}
\end{equation}
where $f_P$ is, by definition,
the pseudoscalar decay constant,
and the sign arises from Wick contractions.
By contrast, the hairpin correlator is observed to
have a leading double-pole contribution:
\begin{equation}
\widetilde\Delta_h(p) = {\CD} \frac{f_P^2}
{(p^2 + M_\pi^2)^2} + {\CS} \frac{f_P^2}{p^2+ M_\pi^2}
+ \textrm{non-pole}
\,,\label{eq:doublepole}
\end{equation}
where $f_P$ is to be determined from eq.~(\ref{eq:singlepole}).
I will refer to the residues $\CD$ and $\CS$ as the hairpin
and single-pole coefficients, respectively.
They are defined to be independent of $p^2$,
but can depend on $m_q$, as do $f_P$ and $M_\pi$.

With this notation in hand, I can now state more precisely
the observations made in BDET and previous work. These are, first,
that $\CS\approx 0$, i.e. that $\widetilde\Delta_h(p)$ is 
consistent with a pure double-pole;
and, second, that $\CD$ has a significant mass dependence.
I now turn to the predictions of the 
quenched chiral Lagrangian for $\CD$ and $\CS$.

\bigskip

To begin with I need to determine the operator 
in the effective theory which matches onto the pseudoscalar density.
This can be done, as in unquenched chiral perturbation theory,
by taking appropriate derivatives with respect to the quark mass matrix,
treating the latter as a source. At leading order, using $\CL_0$, this gives
$\bar q_A \gamma_5 q_B \to 2 i B_0 f \Phi_{AB} + O(\Phi^3)$, 
for all $A$ and $B$, leading to the tree-level results
\begin{eqnarray}
\widetilde\Delta_c(p) &=& - \frac{(2 B_0 f)^2}{p^2 + \chi} 
\\
\widetilde\Delta_h(p) &=& 
\frac{(m_0^2 + \alpha_\Phi p^2)}{3}\frac{(2 B_0 f)^2}{(p^2 + \chi)^2}
\,.
\end{eqnarray}
Comparing to eqs.~(\ref{eq:singlepole}) and (\ref{eq:doublepole})
I find
\begin{equation}
M_\pi^2 = \chi\,; \quad
f_P = 2 B_0 f\,; \quad
\CD = \frac{m_0^2 - \alpha_\Phi M_\pi^2}{3}
\,;\quad
\CS = \frac{\alpha_\Phi}{3}
\,.
\label{eq:treelevel}
\end{equation}
These leading order results are well known~\cite{BG}, and agree with those
quoted in BDET when normalization and sign conventions are taken into account.

As noted above, at this order $\alpha_\Phi$ provides 
both the mass dependence of $\CD$ {\em and}
the value for $\CS$. It is perhaps worth understanding this point
in more detail. One might have thought that a term in $\CL_0$ of the form
\begin{equation}
(f^2/4) v_2 \Str(\chi U^\dagger- U \chi) i \Phi_0/f
\label{eq:v2term}
\end{equation}
would provide additional mass dependence to the hairpin vertex, and thus
break the correlation between $\CD$ and $\CS$. 
However, this term also contributes
to the effective field theory expression for the pseudoscalar density, 
and leads to an additional term in $\CS$, in such a way that $v_2$ can be
absorbed into a redefinition of $\alpha_\Phi$.
This is a consequence of the fact, noted in Ref.~\cite{BG}, that
the $v_2$ term (\ref{eq:v2term}) can be removed by a field
redefinition [explicitly this is $U\to U \exp(-i v_2 \Phi_0/f)$]
and a redefinition of parameters 
($\alpha_\Phi\to \alpha_\Phi - \sqrt3 v_2$).

\bigskip
Before proceeding to the details of the NLO calculation,
it is necessary to discuss the power counting appropriate to quenched
chiral perturbation theory. I give an overview here
and a comprehensive discussion in the appendix.
Standard unquenched chiral perturbation theory is, in the
meson sector, an expansion in 
$\epsilon^2\sim M_\pi^2/\Lambda_\chi^2\sim p^2 /\Lambda_\chi^2$,
where $\Lambda_\chi=4\pi f$.
The issue in quenched chiral perturbation theory is how to treat the
extra constants that appear, the most important of which is $m_0$.
This appears in two dimensionless combinations in the
calculations considered here. First,
loop contributions are suppressed by
powers of $\delta=m_0^2/(3\Lambda_\chi^2)$, which,
as noted above, is numerically small.\footnote{%
In fact, the leading logarithmic corrections are proportional to powers
of $\delta \log(M_\pi/\mu)$, with $\mu$ the regularization scale, and thus
dominate over the leading term in the extreme chiral limit. For most practical
simulations, however, including those of BDET,
the logarithm is of $O(1)$, and does not lead to a significant enhancement.
I assume this to be case in the power counting schemes I use.}
Second, the ratio of the two terms contributing to 
$\CD$ in eq.~(\ref{eq:treelevel}) is $\alpha_\Phi M_\pi^2/m_0^2$.
It is also reasonable to treat this combination as small,
given that $m_0$ is close to $\Lambda_\chi$ and $\alpha_\Phi$ is small.
To do so consistently, however, one must develop a systematic power counting
scheme in which both combinations are small, 
and for which one can assess the relative
size of other contributions.

Two such schemes are explained in the appendix. In the first (the standard or
\textbf{PC1} scheme) one takes
$\delta\sim \alpha_\Phi/3 \sim \epsilon^2$, while in the second
(the \textbf{PC2} scheme) $\delta\sim {\alpha_\Phi/3}\sim \epsilon^{4/3}
$.
In both schemes the $\alpha_\Phi$ contribution to $\CD$ is non-leading:
\begin{equation}
\frac{\alpha_\Phi M_\pi^2}{m_0^2} = \frac{\alpha_\Phi}{3}\frac{\epsilon^2}{\delta}
\sim \epsilon^2 \quad \textbf{PC1/PC2}\,.
\end{equation}
The schemes differ, however, in the size of other contributions.
In the \textbf{PC2} scheme, since $\delta$ is larger than $\epsilon^2$,
``second-order" terms of size $\delta^2$, $\delta\epsilon^2$, and
$\epsilon^4$ are progressively smaller, while all three are treated as of
the same size in \textbf{PC1}. Thus the \textbf{PC2} scheme has some similarities to
that used in BDET, in which terms of relative size 
$\delta \epsilon^2 \log(M_\pi/\mu)$ are kept, but those of size $\epsilon^4$
are not. This point is discussed further in the appendix.

\bigskip

In this paper I calculate the connected and disconnected correlators
at NLO in the \textbf{PC2} scheme. By NLO I mean that corrections
of size up to and including $\epsilon^2$ relative to the leading contribution
are included. After amputation, the disconnected vertex has the form
$\CD + (p^2 + M_\pi^2) \CS + \dots$, from which one can extract $\CD$ and $\CS$.
Note that one obtains a lower order result for $\CS$ than for $\CD$ because of
the explicit factor of $(p^2 + M_\pi^2)$ multiplying $\CS$.

I choose the \textbf{PC2} scheme because it reduces the number of
diagrams that contribute and thus
simplifies the calculation.
In both schemes, however, the general point is the same: there are
other terms which contribute at the same order as the $\alpha_\Phi$ vertex,
and these must be included.

\begin{figure}
\begin{center}
\epsfxsize=\hsize
\epsfbox{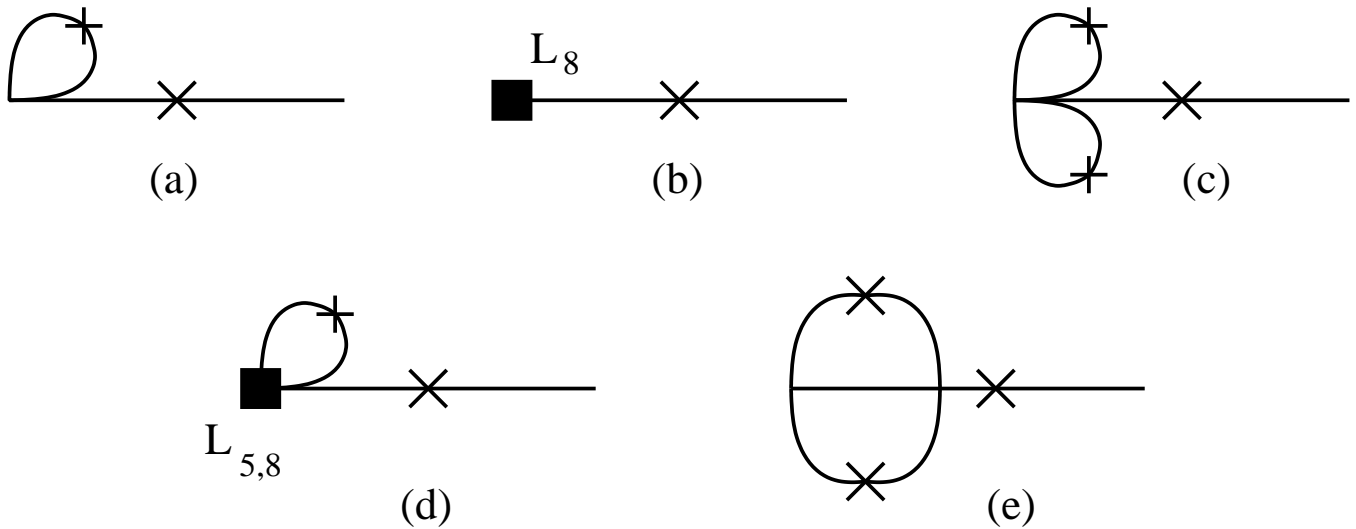}
\end{center}
\caption{Examples of diagrams contributing to 
the pseudoscalar hairpin correlator
which are associated with vertex renormalization.
Simple vertices are from the first two terms in $\CL_0$,
crosses indicate $m_0$ or $\alpha_\Phi$ vertices,
filled boxes are interactions from $\CL_{1A}$, 
with the specific term which contributes being noted.
Note that if the rightmost propagator and hairpin vertex are
removed, these become contributions to the quark-connected
correlator.
}
\label{fig:vertex}
\end{figure}

\begin{figure}
\begin{center}
\epsfxsize=\hsize
\epsfbox{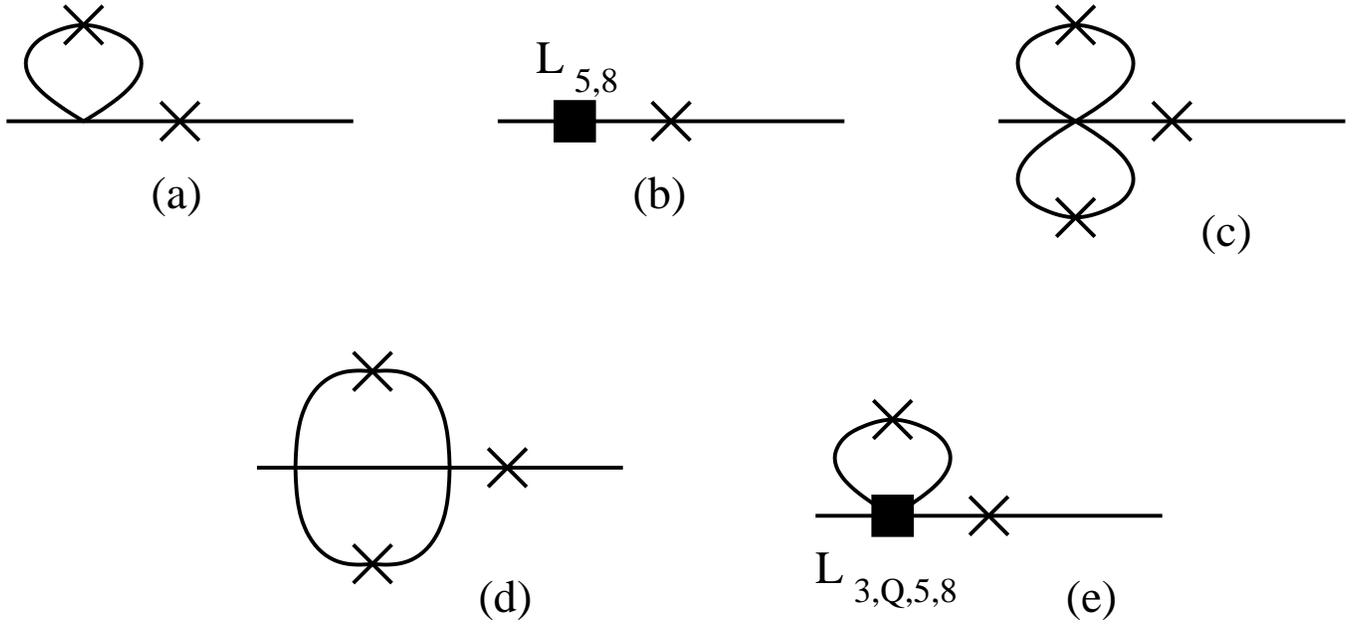}
\end{center}
\caption{Examples of diagrams contributing to 
the pseudoscalar hairpin correlator
which are associated with pion mass and wavefunction
renormalization. Notation is as in fig.~\protect\ref{fig:vertex}.
Note that if the rightmost propagator and hairpin vertex are
removed, these are contributions to the quark-connected
correlator.
}
\label{fig:mass}
\end{figure}

The calculation of the connected correlator is standard
and leads to the replacement of $f_P$ 
and $M_\pi^2$ in eq.~(\ref{eq:singlepole}) with
their NLO expressions. The leading diagrams that contribute
can be deduced from those in
figs.~\ref{fig:vertex} and \ref{fig:mass} 
by removing the rightmost propagator and the hairpin vertex to which
it couples. 
As can be seen from table~\ref{tab:ncut0}, a NLO calculation
involves only diagrams (a) and (b) from both figures---other diagrams are
of higher order than $\epsilon^2$ in both power counting schemes.
The explicit expressions
for $f_P$ and $M_\pi^2$ are given in BDET and elsewhere,
with the latter repeated in the Appendix, eq.~(\ref{eq:MpiNLO}).
They are not needed to determine $\CD$ and $\CS$.

\begin{figure}
\begin{center}
\epsfxsize=\hsize
\epsfbox{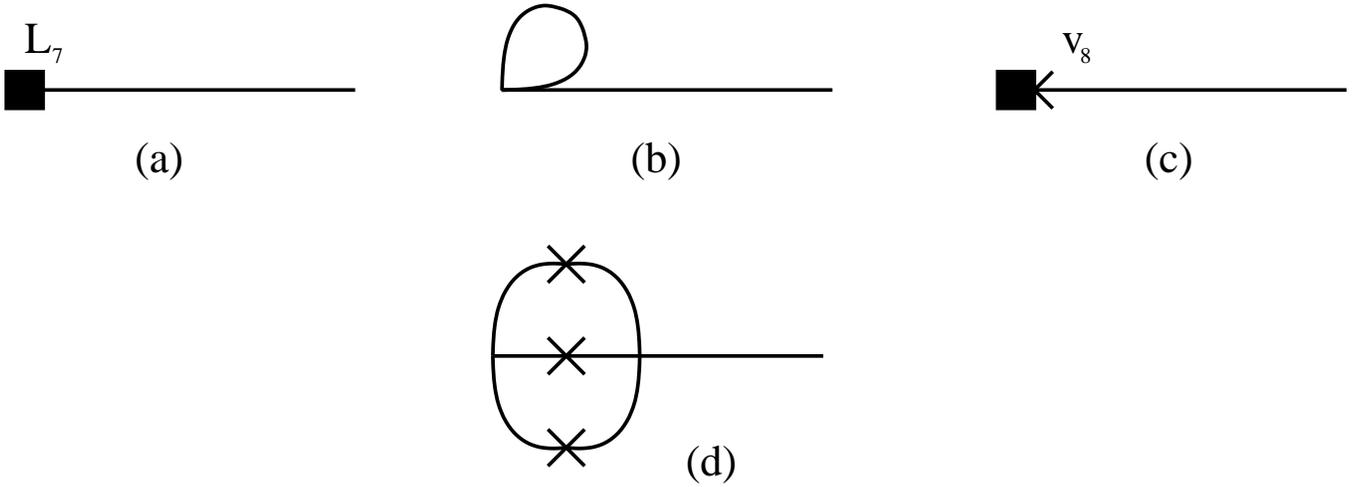}
\end{center}
\caption{Examples of diagrams contributing to 
the pseudoscalar hairpin correlator which contribute
directly to the single pole residue $\CS$.
Notation is as in figs.~\protect\ref{fig:vertex}, 
except that boxes with attached arrowheads are from $\CL_{1B}$
(the arrowhead indicating the $\Phi_0$ field).
}
\label{fig:vertexb}
\end{figure}

\begin{figure}
\begin{center}
\epsfxsize=\hsize
\epsfbox{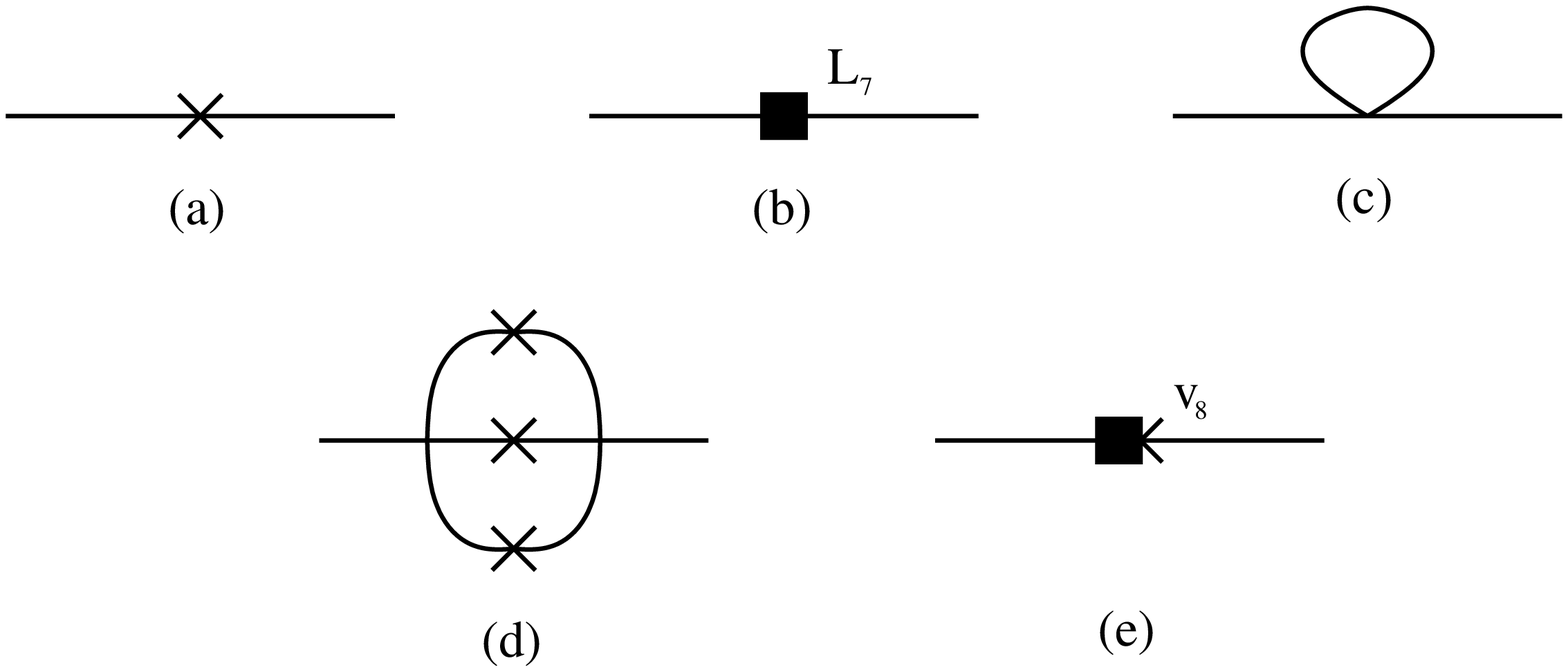}
\end{center}
\caption{Examples of diagrams contributing to 
the pseudoscalar hairpin correlator
which are associated with renormalization of the hairpin vertex
itself.
Notation is as in figs.~\protect\ref{fig:vertex} and
\protect\ref{fig:vertexb}.
}
\label{fig:hairpin}
\end{figure}

For the disconnected correlator, one can divide the contributions
into the four classes.
(1) Those renormalizing
the external pseudoscalar density, and which would be present
also if the density were flavor non-singlet.
Examples of these are shown in fig.~\ref{fig:vertex}.
(2) Those renormalizing the external density, but which are
present only because the density is flavor singlet. These contribute only
to the single pole amplitude $\CS$, but not to $\CD$. Examples
are shown in fig.~\ref{fig:vertexb}.
(3) Those renormalizing the mass and wavefunction of the pseudo-Goldstone
boson, shown in fig.~\ref{fig:mass}.
(4) Those associated with the hairpin vertex, shown in fig.~\ref{fig:hairpin}.
Two observations greatly simplify the calculation.
First, the contributions of classes (1) and (3) are identical
to those appearing in the connected correlator.
This means that the same expressions for $f_P$ and $M_\pi^2$
apply to both connected and disconnected correlators---as one would
have expected.
Second, all contributions of classes (2) and (4), aside from the
leading order diagram fig.~\ref{fig:hairpin}(a),
are smaller than NLO.\footnote{%
This simplification does not occur in the \textbf{PC1} scheme,
in which at NLO one must also calculate
figs.~\ref{fig:vertexb}(a-b) and \ref{fig:hairpin}(b-c).}

These observations imply that the sole NLO correction to the
hairpin vertex in the \textbf{PC2} scheme
comes from wavefunction renormalization. This
in turn arises at NLO only from fig.~\ref{fig:mass}(b),
as fig.~\ref{fig:mass}(a) only renormalizes the mass.
Since there are two 
boson propagators in the disconnected correlator,
compared to only one in the connected correlator, 
only ``half'' of the wavefunction
renormalization in the disconnected case is absorbed into 
the one-loop result for $f_P^2$. The other half
renormalizes the hairpin vertex, leading to the results\footnote{%
In an earlier version of this paper I also kept corrections
of relative size $\epsilon \delta^2 \log(M_\pi/\mu)$, coming from
fig.~\ref{fig:mass}(e), as in the
calculation of BDET. The analysis in the appendix shows that
in either power counting scheme one should then also include
all the other diagrams in figs.~\ref{fig:mass} and \ref{fig:hairpin},
including non-trivial two loop diagrams.
I have not attempted this calculation.}
\begin{equation}
\CD = \frac{1}{3}\left\{m_0^2
\left(1 - 8 L_5 \frac{\chi}{f^2}
%
%
%
\right)- \alpha_\Phi M_\pi^2\right\}
\,;\quad
\CS = \frac{\alpha_\Phi}{3}
\,.
\label{eq:final}
\end{equation}
Note that the NLO contributions do not change $\CS$ from its tree level
value in eq.~(\ref{eq:treelevel}). In particular, the
correction to $\CS$ from wavefunction renormalization
is a correction of relative size $\epsilon^4$  in the amputated disconnected correlator,
and thus beyond the order I am considering.
Note also that
one can interchange $\chi$ and $M_\pi^2$  in the result, as the
difference is of higher order than I consider.

Equation (\ref{eq:final}) is the new result of this paper.
It shows that the mass dependence of the hairpin $\CD$ is {\em not}
tied to the coefficient of the single pole $\CS$. Thus, in particular,
the single-pole term can be absent (as found, to good approximation,
by BDET), but the hairpin vertex can have a linear dependence on $M_\pi^2$.
Furthermore, the latter dependence provides
an alternative method for determining $L_5$.

\bigskip
There is a nice consistency check on the result (\ref{eq:final}).
At $p=0$, the hairpin correlator
is proportional to the 
topological susceptibility of the pure gauge theory~\cite{smitvink}
\begin{equation}
m^2 \widetilde\Delta_h(p=0) = \chi_t = \langle \nu^2\rangle/V 
\,,
\label{eq:topsusc}
\end{equation}
where $m$ is the quark mass, $\nu$ the topological charge,
and $V$ the four dimensional volume.
This is the ``fermionic'' method for measuring the susceptibility.
Since this formula holds for any $m$, and $\chi_t$, being a
pure gauge quantity, is independent of $m$, the left-hand-side
should also be independent of $m$.
Inserting eq.~(\ref{eq:final}) into the general form
(\ref{eq:doublepole}) I find
\begin{eqnarray}
m^2 \widetilde\Delta_h(p=0) &=& 
\frac{m^2 f_P^2}{M_\pi^4}\left(\CD + \CS M_\pi^2 
+ O(M_\pi^4)\right) 
\\
&=&
\frac{m^2 f_P^2 m_0^2}{3 M_\pi^4}
\left(1-8 L_5 \frac{\chi}{f^2} 
\right)
\\
&=& \frac{f_A^2 m_0^2}{12}
\left(1-8 L_5 \frac{\chi}{f^2} 
\right)
\,.
\end{eqnarray}
where in the last step I have used the axial Ward identity
$2 m f_P = M_\pi^2 f_A$, which holds to all orders in chiral
perturbation theory (and can be checked at NLO from the expressions
in BDET).\footnote{%
Here I use the BDET normalization such that $f_A=\sqrt2 f$ in the
chiral limit.}
Using the expression given in BDET for the axial decay constant $f_A$,
I find
\begin{equation}
m^2 \widetilde\Delta_h(p=0) = \frac{f^2 m_0^2}{6}
\,.
\label{eq:topsuscf}
\end{equation}
This is independent of the quark mass, as required.

One can try turning this argument around, i.e.
assuming (\ref{eq:topsuscf}) and deriving eq.~(\ref{eq:final}).
This will, however, only constrain the combination $\CD + M_\pi^2 \CS$,
and not allow a separate determination of $\CD$ and $\CS$.

\bigskip
I close this section with a technical comment. 
In order to predict the coefficient of the
single pole in the hairpin correlator one must control
the corrections to the operator which creates the meson
at $O(p^2)$ relative to the leading order. This is because
such a correction multiplying the double-pole 
leads to a contribution to the single-pole.
Thus it is crucial that one calculate the hairpin correlator
with an operator which can be matched onto the effective
theory without introducing additional coefficients at $O(p^2)$.
This is only true for the pseudoscalar density, used here,
and for the axial current.
With any other operator coupling to the pseudoscalar meson,
e.g. $\bar q \partial^2 \gamma_5 q$, the coefficient of
the single pole could not be predicted.

In light of this discussion, one might be
concerned about the need to include the operator
\begin{equation}
\Str\left(\chi \partial^2 [U + U^\dagger]\right) 
\end{equation}
in $\CL_{1A}$, since this would give a contribution to the pseudoscalar
density proportional to $p^2 \Phi$.
It turns out, however, that this operator can be removed
(as in unquenched chiral perturbation theory) by a field definition
and appropriate shifts in $L_5$, $L_8$ and $\alpha_\Phi$.
Thus it is a redundant operator.\footnote{%
I am grateful to Bill Bardeen for helpful discussions on this point.}
I have checked this by explicitly calculating its contributions
to $f_A$, $f_P$ and $\widetilde\Delta_{c,h}$.

\section{Conclusion}
\label{sec:conc}

In this note I have presented a complete NLO
calculation of the hairpin vertex in quenched chiral perturbation theory.
The result, eq.~(\ref{eq:final}),
resolves a puzzle concerning the determination of
the parameter $\alpha_\Phi$. The method I had proposed in 
Ref.~\cite{SSLAT96}, using the quark mass dependence of the 
residue of the double-pole, turns out to be incorrect.
The correct method is to use the residue of the single-pole,
which, according to BDET, leads to $0 \approx \alpha_\Phi \ll 1$.
The mass dependence of the double-pole residue then allows
one to determine $L_5$. 

It will be interesting to reanalyze
the results of BDET and earlier work in this light,
and compare the determinations of $L_5$ with those obtained
from the mass dependence of the axial decay constant, $f_A$.
As a rough first attempt, I note that my earlier estimate 
$\alpha_\Phi\approx 0.5$ converts to
$L_5 \approx 0.5/(48 \pi^2 \delta) \approx 1.3 \times 10^{-3}$,
where I have taken $\delta=0.1$ from BDET for the value 
of the hairpin vertex in the chiral limit.
This is close to the values for $L_5$ quoted in BDET.

The calculation also suggests a simple way of displaying the
consistency of the different determinations of the parameters
of quenched chiral perturbation theory. 
It follows from the previous considerations that 
\begin{equation}
(\CD + M_\pi^2 \CS) f_A^2 = 4 \chi_t + O(M_\pi^4)
\,.
\label{eq:consistency}
\end{equation}
In the notation of BDET, this means that the combination
$[(m_0^{\rm eff})^2 + M_\pi^2 \alpha_\Phi]\times f_A^2$
should be independent of $m$ at linear order.
Since $\alpha_\Phi$ is small,
this constraint explains the observed result that
$m_0^{\rm eff}$ decreases with increasing $m$
in terms of the fact that $f_A$ increases with $m$.

When applying the results of this paper to numerical simulations,
one should keep in mind that the previous discussion did not
account for finite volume or discretization errors.
Concerning the former, one can include the leading effect by replacing
the loop integrals with those appropriate to finite volume.
As for the discretization errors, if one uses
Wilson-like fermions, then there will be $O(am)$
corrections to the results presented here, unless the action and operators
are non-perturbatively $O(a)$ improved (which is not the
case in the work of BDET).
This is because of
the breaking of chiral symmetry by the action.
In particular, the  quantity in eq.~(\ref{eq:consistency})
will have a mass dependence proportional to $am$.
Thus a possible use of this relation is to gauge the size
of such discretization errors.

\section*{Acknowledgments}
I thank Martin Savage and Ruth Van de Water
for comments on the manuscript,
and Bill Bardeen and Maarten Golterman for discussions.
This work was supported in part by the US Department of Energy through grant
DE-FG03-96ER40956/A006.

\appendix
\section{Power counting in quenched chiral perturbation theory}

In this appendix I discuss power
counting schemes for quenched chiral perturbation theory,
and explain that used for the calculation described
in the main text.
I also take the opportunity to collect
some results that are presumably known to workers in the field
but which, as far as I know, are not available in the literature.
I use the standard notation for the chiral power counting parameter:
$\epsilon^2 = \chi/\Lambda_\chi^2$,
where $\chi$ refers to either $M_\pi^2$ or $p^2$.

To make the arguments I will frequently use the
fact that, in the quenched theory,
diagrams involving pions
(and their graded partners) can be converted
in a precise, well-defined way, into those
involving quark (and ghost quark) lines.
This involves following flavor indices, and is allowed 
because of the necessary inclusion of the $\Phi_0$ field
in quenched chiral perturbation theory~\cite{BG,ShShII}.
The advantage of the quark-line representation is that the
impact of quenching is transparent:
all diagrams with internal quark lines cancel against
those with corresponding ghost quark lines.
In the following I only consider the diagrams which remain
after such cancellations have occurred,
i.e. I consider power counting
only for diagrams that give non-trivial contributions.

\subsection{Power counting with $\CL_0$}

The largest contributions to correlators involve vertices from $\CL_0$,
eq.~(\ref{eq:L0}),
so I consider first the power counting appropriate for this Lagrangian alone.
For diagrams with no $m_0^2$ and $\alpha_\Phi$ vertices,
 standard chiral power counting gives the following result
for the amputated N-point correlation function with external
momenta of $O(M_\pi^2)$:%
\footnote{
This also holds for the corresponding 1PI correlator,
and thus for the N-point scattering amplitude,
although I will not use this here.}
\begin{equation}
\CA_N \sim \frac{\chi}{f^{N-2}} \left({\epsilon^2}\right)^{n_\ell}
\,, \qquad
n_\ell = \mathrm{number\ of\ loops}
\,.
\label{eq:QCDCL0count}
\end{equation}
For example, at leading order 
the two-point correlator gives the pion mass,
$\CA_2 \sim \chi \sim M_\pi^2$, with corrections suppressed
by the number of loops.

This result, and those that follow, also apply to the N-point
correlator of the pseudoscalar density, except that
eq.~(\ref{eq:QCDCL0count}) must be multiplied by
$f^N$. Since what matters is the relative size of different
terms, and not the overall factor, the ensuing analysis will
be directly applicable to the correlators considered in
the main text.

Including $m_0^2$ and $\alpha_\Phi$ vertices, I note that each such vertex
introduces an extra propagator compared to the diagram without
the vertex, and so their contributions (after subtracting
divergences so that loop momenta are of order $M_\pi^2$) are
\begin{equation}
m_0^2 \ \mathrm{vertex} \sim 
\frac{m_0^2/3}{p^2 + M_\pi^2}\sim \frac{m_0^2}{3\chi} 
=  \frac{\delta}{\epsilon^2}
\,,\qquad
\alpha_\Phi \ \mathrm{vertex} \sim 
\frac{p^2 \alpha_\Phi/3}{p^2 + M_\pi^2}\sim \alpha_\Phi/3
\,.
\label{eq:deltasize}
\end{equation}
I include the factors of $1/3$ since,
given the definition $\Phi_0=\Str(\Phi)/\sqrt{3}$,
they are generic.
Including these results the power counting for $\CL_0$ is
\begin{equation}
\CA_N \sim \frac{\chi}{f^{N-2}} \left({\epsilon^2}\right)^{n_\ell- n_\delta}
(\delta)^{n_\delta} 
\left(\frac{\alpha_\Phi}{3}\right)^{n_\alpha}
\,,
\label{eq:L0count}
\end{equation}
where $n_\delta$ and $n_\alpha$ are, respectively,
the number of $m_0^2$ and $\alpha_\Phi$ vertices.
The appearance of $1/\epsilon^2$ in the contribution
of the $m_0$ vertex in eq.~(\ref{eq:deltasize})
is the reason for the modification of power counting in
quenched chiral perturbation theory.
As discussed below, one typically 
chooses $\delta \ge \epsilon^2$, so that
additional $m_0^2$ vertices either come at ``no cost''
or are enhanced.

There is, however, a constraint on the number of $m_0^2$ and $\alpha_\Phi$
vertices due to the cancellation of internal quark and ghost loops.
A simple example of this constraint
is that one cannot have two such
vertices on the same propagator, since this would require
an internal quark loop.
As I will explain, the general form of this constraint is
\begin{equation}
n_\delta + n_\alpha \le n_\ell + n_{cut}
\,,
\label{eq:L0const}
\end{equation}
where $n_{cut}$ is the number of independent purely gluonic
``cuts" through the correlator.
In other words, if one traces the quark lines, they will fall into
$n_{cut}+1$ quark-disconnected components. In perturbation theory
the only connection between these components is provided by
gluons. In the cases of interest here,
the connected correlator $\widetilde\Delta_c(p)$ has $n_{cut}=0$,
while the hairpin correlator $\widetilde\Delta_h(p)$ has $n_{cut}=1$.

To demonstrate the result (\ref{eq:L0const}) I first consider
correlators with $n_{cut}=0$, i.e. those 
which are connected in terms of quark lines.
I need the result that the four and higher-point
vertices coming from $\CL_0$ involve a single
supertrace, and thus are themselves singly connected
in terms of quark line diagrams.
The cancellation of diagrams with internal quark loops then implies
the following: there must be at least one path through the
(pion language) diagram from each $m_0^2$ or $\alpha_\Phi$ vertex
to each of the external pions which does not pass through another
$m_0^2$ or $\alpha_\Phi$ vertex. This must be true if one leaves the original
vertex in either direction.
Were this not the case, there would
be a segment of the diagram completely ``cut-off'' by $m_0^2$ and
$\alpha_\Phi$ vertices, corresponding to an internal quark loop
in the quark-line language.

A corollary of this result is that, if one were to ``pull apart"
each of the $m_0^2$ and $\alpha_\Phi$ vertices to give two extra
external pions for each such vertex, then the resulting ``expanded"
pion-language diagram is singly connected (in the usual sense).

Now imagine starting with an unquenched chiral perturbation
theory diagram and adding as many $m_0^2$ and $\alpha_\Phi$ vertices
as possible on the propagators. When this has been done, the
corresponding expanded diagram must be a tree diagram,
since if there were any loops, the internal propagator(s) could
be expanded by adding additional $m_0^2$ or $\alpha_\Phi$ vertices.
It must also be a singly connected (pion-language) diagram from
the corollary of the previous paragraph.
Now, when one now sews the $m_0^2$ and $\alpha_\Phi$ 
vertices back together again,
each leads to an independent loop momentum 
in the original diagram,
and thus $n_\delta+n_\alpha=n_\ell$.
Since this is the maximum number of such vertices that
can be added, one finds the inequality (\ref{eq:L0const})
with $n_{cut}=0$.

The generalization to $n_{cut}>0$ is straightforward.
One can now have additional $m_0^2$ and/or $\alpha_\Phi$ vertices,
since $n_{cut}$ of them can be used to separate the $n_{cut}+1$
quark-disconnected components of the diagram. Thus one finds
the general inequality of eq.~(\ref{eq:L0const}).

When using this formula two additional results should be borne in mind.
First, there is, in general,
no non-trivial lower limit on the
number of $m_0^2$ and $\alpha_\Phi$ vertices.
One might have expected the constraint
$n_\delta + n_\alpha \ge n_{cut}$, i.e. that
$m_0^2$ and $\alpha_\Phi$ vertices are necessary to
lead to quark-disconnected components. This is not the case,
however, because loop diagrams involving connected vertices
can lead to quark-disconnected correlators, as discussed
in Refs.~\cite{SSQChlog0,ShShI} for the hairpin vertex.
Second, even if a given choice
of $n_\delta$, $n_\alpha$ and $n_\ell$ satisfies the
constraint in eq.~(\ref{eq:L0const}), there may not
be a diagram contributing to $\CA_N$. Examples of this
will be seen below.

\subsection{Including $\CL_{1}$ and higher order Lagrangians}

It is straightforward to include the ``standard'' $O(\epsilon^4)$ Lagrangian,
$\CL_{1A}$, in the power counting.
Each vertex from $\CL_{1A}$ gives an
additional factor of $\epsilon^2$,
as in the unquenched theory, and
the constraint on $n_\delta+n_\alpha$ remains unchanged.
It should be borne in mind, however, that since some of the vertices
in $\CL_{1A}$ contain two supertraces, it becomes more complicated
to determine the ``quark-line connectivity'' of a given graph.
This has no impact on the generic power counting, but means that a
larger number of diagrams will vanish in the quenched theory.

As noted in the text, one must also consider operators which
are built upon those in $\CL_0$ and $\CL_{1A}$ by multiplying
by additional factors of $\Phi_0/f$. 
Two examples are given in eq.~(\ref{eq:L1B}).
The key observation is that these factors have 
the same power of $1/f$ as vertices
arising from the expansion of $U=\exp(2i\Phi/f)$.
Thus the power counting is unchanged, as long as $n_\delta$
and $n_\alpha$ are
generalized to be the number of vertices built upon
the original $m_0^2$ and $\alpha_\Phi$ vertices, respectively.
Thus, for example, $n_\delta$ becomes the 
total number of vertices of the form $(\Phi_0)^q$ with $q=2,4,6,\dots$.
What is changed is the connectivity of the diagram.
Each extra factor of $\Phi_0/f$ requires an
additional loop, or gluonic cut, in the diagram,
since $\Phi_0$ closes off quark lines. Thus the loop
constraint generalizes to
$n_\delta + n_\alpha + n_\phi \le n_\ell + n_{cut}$,
where $n_\phi$ is the total number of extra factors of $\Phi_0$
in the diagram.
Note that a $(\Phi_0)^4$ vertex contributes both to
$n_\delta$ (which it increments by one) and to
$n_\phi$ (which it increments by two).

Each vertex involving extra factors of $\Phi_0$ comes with
a new coupling, e.g. $v_{0,8}$ in $\CL_{1B}$.
These couplings are suppressed in the large $N_c$ limit
because a $\Phi_0$ vertex is connected only by gluons to
the rest of the diagram. Each additional $\Phi_0/f$ leads
to a suppression by $1/N_c$.
For example, if the quark loop is connected by two gluons to the rest
of the diagram, there is an additional color loop, and the counting is
$g^4 N_c=(g^2 N_c)^2/N_c \sim 1/N_c$. 
(The $\sqrt{N_c}$ contained in $f$ is exactly that needed to
create a state which is correctly normalized in the large $N_c$ limit.)
Thus I find that $v_8\sim 1/N_c$ and $v_0\sim 1/N_c^2$.
For a general diagram, the additional suppression is by $(N_c)^{-n_\phi}$.

I should stress that I am not claiming to
extract the full $N_c$ dependence of diagrams.
There is also implicit $N_c$ dependence in $f$, 
in some of the Gasser-Leutwyler
coefficients, $L_i$, and in $m_0^2$ and $\alpha_\Phi$.
Indeed the latter two are both proportional to $1/N_c$~\cite{BG},
My idea is that, based on the general success of the OZI rule,
it is likely that constants such as $v_8$ and $v_0$ are suppressed
relative to unity. Keeping track of the powers of $1/N_c$ associated
with these constants allows one to implement this as one deems
appropriate. 

Higher order terms in the quenched chiral Lagrangian involve
more derivatives (acting on either $U$ or $\Phi_0$) and
more factors of $\chi$.
These can be accounted for, along with those
from vertices in $\CL_{1A}$, by introducing the quantity $n_1$,
which counts the number of factors of $\chi$  plus
half the number of derivatives 
{\em beyond those that would appear if the vertices were
from the lowest order Lagrangian $\CL_0$}.
Thus a vertex from the standard chiral Lagrangian with six derivatives
and one factor of $\chi$ would increment $n_1$ by three.
Overall, 
diagrams are multiplied by an extra factor of $(\epsilon^2)^{n_1}$.

There is an ambiguity in this definition of $n_1$: does, for
example, the vertex proportional to $(\partial \Phi_0)^4$ count
as $n_\delta=1$, $n_\phi=2$ and $n_1=2$ or as
$n_\alpha=1$, $n_\phi=2$ and $n_1=1$?
In other words, does it contribute a correction proportional to 
$\delta\epsilon^2/N_c^{2}$ or to $(\alpha_\Phi/3) \epsilon^4/N_c^2$?
These are not the same size in some power counting schemes,
e.g. those discussed below.
The ambiguity arises because dimensions can be balanced either
by powers of $m_0$ or powers of $f$.
In fact, this ambiguity also arises
for the $\Phi_0^q$ vertices without derivatives, and I have chosen
there the overall factor of $m_0^2/f^{q-2}$ rather than $1/f^{q-4}$.
Fortunately, for the calculation
considered in the main text, I can leave this ambiguity
unresolved,
as the operators in question do not contribute until higher
order than that I consider.

The preceding considerations together imply the following
final power counting result:
\begin{equation}
\CA_N \sim \frac{\chi}{f^{N-2}} 
\left({\epsilon^2}\right)^{n_\ell- n_\delta + n_{1}}
(\delta)^{n_\delta} 
\left(\frac{\alpha_\Phi}{3}\right)^{n_\alpha}
\left(\frac{1}{N_c}\right)^{n_\phi}
\,, \qquad
n_\delta + n_\alpha + n_\phi \le n_\ell + n_{cut}
\,.
\label{eq:L01count}
\end{equation}

\subsection{Applying the power counting}

I now apply the result (\ref{eq:L01count}) to the examples
considered in the text, namely the two-point quark-connected
and disconnected correlators (i.e. $\CA_2$ with
$n_{cut}=0$ and $1$ respectively).
To do so I must decide how to treat $\delta$, $\alpha_\Phi$
and $1/N_c$ relative to each other and to $\epsilon^2$. This
depends, of course, on the size of quark masses considered.
I have in mind mesons somewhat lighter than the kaon mass,
so that $\epsilon^2\approx 0.2$.

I consider two possibilities for the relative size of
the different terms.
The first is
\begin{equation}
\mathbf{PC1:}\qquad
\delta \sim \frac{\alpha}{3} \sim \frac{1}{N_c^2} \sim \epsilon^2
\qquad\Rightarrow\quad
\CA_N \sim \frac{\chi}{f^{N-2}} 
\left({\epsilon^2}\right)^{n_\ell + n_{1} + n_{\alpha} + n_\phi/2}
\,,
\end{equation}
which one might call ``standard" power counting.
The key choices here are that $\delta\sim\epsilon^2$, so that there
is no extra enhancement of the $m_0^2$ vertices,
and that $\alpha_\Phi/3$ is small. One might argue that 
$\alpha_\Phi/3\sim \epsilon$, rather than $\epsilon^2$, because
it is proportional to $1/N_c$. I prefer the smaller choice given
the numerical evidence that $\alpha_\Phi\ll 1$.

The second scheme is
\begin{equation}
\mathbf{PC2:}\qquad
\delta \sim \frac{\alpha}{3} \sim \frac{1}{N_c^2} \sim \epsilon^{4/3}
\qquad
\CA_N \sim \frac{\chi}{f^{N-2}} 
\left({\epsilon^2}\right)^{n_\ell + n_{1} 
-n_\delta/3+ 2n_{\alpha}/3 + n_\phi/3}
\,.
\end{equation}
This enhances the contributions from $\Phi_0$ vertices relative
to ``standard'' chiral vertices, and is the power counting used
in the text. The power of $\epsilon$
is chosen for a reason that will be explained below. In fact, 
most features of the power counting,
including the application discussed in the main text,
are unchanged
if $\epsilon^{4/3}$ is replaced with $\epsilon$.

Table~\ref{tab:ncut0} collects
the contributions to the quark-connected correlator which are
of size $\epsilon^4$ or smaller relative to
the leading order, in one or both power counting schemes.
Table~\ref{tab:ncut1} gives the analogous contributions to
the hairpin correlator, though only up to size $\epsilon^{10/3}$.
I list all contributions consistent with the constraint
in eq.~(\ref{eq:L01count}), but, as can be seen from the table,
some are absent in quenched chiral perturbation theory.
For example, the standard one-loop ``tadpole' diagram
($n_\ell=1$, all other $n$'s zero) requires an internal quark
loop for $n_{cut}=0$, although, as observed in
Refs.~\cite{SSQChlog0,ShShI}, it is present for $n_{cut}=1$.
Another constraint is that the only term with $n_\phi=1$
and $n_1=0$, namely the interaction quoted in eq.~(\ref{eq:v2term}),
can be (and I assume has been) removed by a field redefinition.

\begin{table}[tb]
\caption{Leading contributions to the quark-connected
two-point correlator in quenched chiral perturbation theory.
\textbf{PC} gives the size of the diagram 
relative to $\CA_2=\chi$ using
the general power-counting formula, 
eq.~(\protect\ref{eq:L01count}), while \textbf{PC1}
and \textbf{PC2} refer to the two specific schemes discussed
in the text. All contributions up to relative size $\epsilon^4$ in
one or both power-counting schemes are included.
The references to fig.~\protect\ref{fig:mass} refer to the parts of
the diagrams to the left of the rightmost hairpin vertex
(which are thus quark-connected).}
\label{tab:ncut0}
\begin{center}
\begin{tabular}{c c c c c c c c c c}
$n_\ell$& $n_\delta$& $n_\alpha$& $n_1$& $n_\phi$ &
\textbf{PC} & \textbf{PC1} & \textbf{PC2} & Figure & Comment
\\ \hline\hline
0 &0& 0& 0& 0& 1& 1& 1&  & \\
0 &0& 0& 1& 0& $\epsilon^2$ & $\epsilon^2$ & $\epsilon^2$ 
   & \ref{fig:mass}(b) &$L_5$ and $L_8$ contribute  \\
0 &0& 0& 2& 0& $\epsilon^4$ & $\epsilon^4$ & $\epsilon^4$ & & \\
\hline
1 &0& 0& 0& 0& $\epsilon^2$  &  & &  & Absent\\
1 &1& 0& 0& 0& $\delta$ & $\epsilon^2$ & $\epsilon^{4/3}$ 
   & \ref{fig:mass}(a)	& Quenched Chiral Log \\
1 &0& 0& 0& 1& $\epsilon^2/N_c$ & & &  & Absent\\
1 &0& 1& 0& 0& $(\alpha_\Phi/3)\epsilon^2$ 
   & $\epsilon^4$ & $\epsilon^{10/3}$ 
      & \ref{fig:mass}(a) & Leading $\alpha_\Phi$ term \\
1 &1& 0& 1& 0& $\delta\epsilon^2$ & $\epsilon^4$ & $\epsilon^{10/3}$ 
   & \ref{fig:mass}(e) & $L_3$, $L_5$, $L_8$, and $L_Q$ contribute\\
1 &0& 0& 1& 0& $\epsilon^4$ & $\epsilon^4$ & $\epsilon^4$ 
   &  &\\
\hline
2 &1& 0& 0& 0& $\delta\epsilon^2$  & & & & Absent\\
2 &2& 0& 0& 0& $\delta^2$ & $\epsilon^4$ & $\epsilon^{8/3}$ 
   & \ref{fig:mass}(c,d) & Full result not available\\
2 &0& 0& 0& 0& $\epsilon^4$ & $\epsilon^4$ & $\epsilon^4$ 
   & & \\
2 &1& 0& 0& 1& $\delta\epsilon^2/N_c$ & & & & Absent\\
\hline
3 &3& 0& 0& 0& $\delta^3$ & $\epsilon^6$ & $\epsilon^{4}$ & &\\
\hline\hline

\end{tabular}
\end{center}
\end{table}

\begin{table}[tb]
\caption{Leading contributions to the hairpin
(quark-disconnected)
two-point correlator in quenched chiral perturbation theory.
All contributions up to $\epsilon^{10/3}$ relative to the
leading order term in
one or both power-counting schemes are included.
Notation as in Table~\protect\ref{tab:ncut1}.
The references to figs.~\protect\ref{fig:mass} 
and \protect\ref{fig:hairpin} refer to the entire diagram.}
\label{tab:ncut1}
\begin{center}
\begin{tabular}{c c c c c c c c c c}
$n_\ell$& $n_\delta$& $n_\alpha$& $n_1$& $n_\phi$ &
\textbf{PC} & \textbf{PC1} & \textbf{PC2} & Figure & Comment
\\ \hline\hline
0 &0& 0& 0& 0& 1  & & &  & Absent\\
0 &1& 0& 0& 0& $\delta/\epsilon^2$ & 1 & $\epsilon^{-2/3}$ 
   & \ref{fig:hairpin}(a) & $m_0$ vertex \\
0 &0& 1& 0& 0& $(\alpha_\Phi/3)$ & $\epsilon^2$ & $\epsilon^{4/3}$ 
   & \ref{fig:hairpin}(a) & $\alpha_\Phi$ vertex \\
0 &1& 0& 1& 0& $\delta$ & $\epsilon^2$ & $\epsilon^{4/3}$ 
   & \ref{fig:mass}(b) & Wavefunction renormalization from $L_5$ \\
0 &0& 0& 1& 0& $\epsilon^2$ & $\epsilon^2$ & $\epsilon^2$ 
   & \ref{fig:hairpin}(b) & Two-supertrace vertex from $L_7$ \\
0 &0& 0& 0& 1& $1/N_c^2$ & & & & Absent \\
0 &0& 0& 1& 1& $\epsilon^2/N_C$ & $\epsilon^3$ & $\epsilon^{8/3}$ 
   & \ref{fig:hairpin}(e) & $v_8$ vertex \\
\hline
1 &0& 0& 0& 0& $\epsilon^2$ & $\epsilon^2$ & $\epsilon^2$ 
   & \ref{fig:hairpin}(c) & Present even in quenched theory\\
1 &1& 0& 0& 0& $\delta$ & & & & Absent \\
1 &2& 0& 0& 1& $\delta^2/\epsilon^2$ & $\epsilon^2$ & $\epsilon^{2/3}$ 
   & \ref{fig:mass}(a)		& No contribution to $\CD$ or $\CS$ \\
1 &2& 0& 1& 0& $\delta^2$ & $\epsilon^4$ & $\epsilon^{8/3}$ 
 & \ref{fig:mass}(e) & Wavefunction renormalization from $L_3$, $L_5$ and $L_Q$ \\
1 &1& 1& 0& 0& $\delta(\alpha_\Phi/3)$ & $\epsilon^4$ & $\epsilon^{8/3}$ 
   & \ref{fig:mass}(a)		& No contribution to $\CD$ or $\CS$ \\
1 &1& 0& 0& 1& $\delta/N_c$ & & & & Absent\\
\hline
2 &3& 0& 0& 0& $\delta^3/\epsilon^2$ & $\epsilon^4$ & $\epsilon^2$ 
   & \ref{fig:mass}(c,d),\ref{fig:hairpin}(d)  & Full result not yet available \\
2 &2& 0& 0& 0& $\delta^2$ & $\epsilon^4$ & $\epsilon^{8/3}$ &  &\\
\hline\hline

\end{tabular}
\end{center}
\end{table}

To illustrate the difference between the power counting
schemes, and the general nature of the chiral expansion in
quenched chiral perturbation theory, I discuss the
application to the pseudo-Goldstone pion mass.
This can be obtained from the amputated quark-connected vertex.
In the \textbf{PC1} scheme, NLO contributions to
the quark-connected correlator are from the Gasser-Leutwyler constants
(entering at tree level) and the $m_0^2$ vertex (at one loop).
The result for the pion mass is
\begin{equation}
\frac{M_\pi^2}{\chi} = 1 - 
\delta[\log(\chi/\mu^2)-1]  +8(2L_8-L_5) \frac{\chi}{f^2}
\,,
\label{eq:MpiNLO}
\end{equation}
in dimensional regularization and the renormalization scheme of Ref.~\cite{DGH}.
The $\alpha_\Phi$ vertex first contributes at NNLO, and is but
one of many terms, including three-loop diagrams with three
$m_0^2$ vertices!

The {\bf PC2} scheme is in part an attempt to systematize the power
counting scheme used in BDET. This involves keeping corrections of
size $\delta \epsilon^2 \log(M_\pi)$,
while those of size $\epsilon^4$ are dropped.
Here I am not treating $\log(M_\pi/\mu)$ as large, so I must
take $\delta$ larger than $\epsilon^2$, which leads to a scheme
like \textbf{PC2}. The choice $\delta\sim\epsilon^{4/3}$ is
made (as opposed to $\delta\sim\epsilon$, say)
so that the three loop $\delta^3$ contribution is smaller
than the $\delta \epsilon^2$ correction. Nevertheless
Table~\ref{tab:ncut0} shows that keeping the $\delta\epsilon^2$
correction, which is of size $\epsilon^{10/3}$, requires keeping
many other terms. Most have been calculated, but not all.
Those not yet calculated are the  non-trivial two-loop diagrams involving
two $m_0^2$ vertices [Fig.~\ref{fig:mass}(d)],
and a subset of the contributions of size $\sim\delta\epsilon^2$
from Fig.~\ref{fig:mass}(e). The latter receives
contributions not only from the $L_5$ and $L_8$ vertices
(calculated in BDET) but also from the $L_3$ and $L_Q$ vertices
(not yet calculated).

One can go to higher order in $\delta$ if one assumes that
$\log(M_\pi/\mu)$ is large, and keeps all terms
of size $[\delta\log(M_\pi/\mu)]^n$. This requires
calculating only the tadpole loops, the simplest
examples being figs.~\ref{fig:mass}(a) and (g),
and these can be summed to all orders~\cite{SSChlog}.
This may be required when working
at the smallest quark masses used in present quenched
simulations~\cite{Liu,Chiu,Gattringer}.

The results in Table~\ref{tab:ncut1} are used in the
calculation in the text, and are discussed further there.
I only note here that an important simplification is that
fig.~\ref{fig:mass}(a) does not contribute to $\CD$ and $\CS$. 
This would have been the leading correction to $\CD$ and the leading
term in $\CS$. It does not contribute because the tadpole loops involving
the $m_0^2$ vertex do not give rise to wavefunction
renormalization. This follows in turn because the vertex coupling
two $\Phi_0$ fields to two non-singlet pions comes only from the
mass term $\Str(\chi U)+ h.c.$ and not from the derivative term 
$\Str(\partial U\partial U)$.\footnote{%
This simplification does not
hold for the four derivative single supertrace vertices
with couplings $L_3$ and $L_Q$, which contribute through
fig.~\ref{fig:mass}(e), as can be seen in the table.
}

\end{document}